\documentclass[10pt,superscriptaddress,twocolumn,amsmath,amssymb,aps,prb,reprint]{revtex4-2}

\makeatletter
\renewcommand{\frontmatter@footnote@produce}[1]{\@empty}
\makeatother

\usepackage{times}
\usepackage{amsfonts}
\usepackage{mathrsfs}
\usepackage[version=3]{mhchem}
\usepackage{graphicx}
\usepackage{dcolumn}
\usepackage{bm}
\usepackage{color}

\usepackage[colorlinks,bookmarks=false,citecolor=blue,linkcolor=red,urlcolor=blue]{hyperref}
\allowdisplaybreaks

\usepackage{physics}
\usepackage{tikz-feynman}
\usepackage{tikz,tikz-feynhand} 
\usepackage{subfigure}
\usepackage{float} 

\def\be{\begin{equation}}       \def\ee{\end{equation}}
\def\bea{\begin{eqnarray}}      \def\eea{\end{eqnarray}}

\begin{document}

\title{Detecting pairing symmetry of bilayer nickelates using electronic Raman scattering}

\author{Jun Zhan}
\affiliation{Beijing National Laboratory for Condensed Matter Physics and Institute of Physics, Chinese Academy of Sciences, Beijing 100190, China}
\affiliation{School of Physical Sciences, University of Chinese Academy of Sciences, Beijing 100049, China}

\author{Matías Bejas}
\affiliation{Facultad de Ciencias Exactas, Ingeniería y Agrimensura and Instituto de Física Rosario (UNR-CONICET),
Avenida Pellegrini 250, 2000 Rosario, Argentina}

\author{Andreas P. Schnyder}
\affiliation{Max Planck Institute for Solid State Research, Heisenbergstrasse 1, D-70569 Stuttgart, Germany}
\author{Andrés Greco}
\affiliation{Facultad de Ciencias Exactas, Ingeniería y Agrimensura and Instituto de Física Rosario (UNR-CONICET),
Avenida Pellegrini 250, 2000 Rosario, Argentina}

\author{Xianxin Wu}\email{xxwu@itp.ac.cn}
\affiliation{Institute for Theoretical Physics, Chinese Academy of Sciences, Beijing, China}

\author{Jiangping Hu}\email{jphu@iphy.ac.cn}
\affiliation{Beijing National Laboratory for Condensed Matter Physics and Institute of Physics, Chinese Academy of Sciences, Beijing 100190, China}
\affiliation{Kavli Institute for Theoretical Sciences, University of Chinese Academy of Sciences, Beijing 100190, China}
\affiliation{ New Cornerstone Science Laboratory, Beijing 100190, China}


	



\date{\today}

\begin{abstract}

The recent discovery of high-temperature superconductivity in both bulk and thin-film bilayer nickelates La$_3$Ni$_2$O$_7$ has garnered significant attention. However, the corresponding pairing symmetry remains debated in both experiments and theoretical studies due to conflicting experimental evidence from bulk and thin-film materials. In this work, we examine the electronic Raman response across different channels for various pairing symmetries within a two-orbital bilayer model. By comparing Raman susceptibilities obtained from multiorbital and band-additive approaches, we demonstrate that Raman response can distinguish between different pairing symmetries and identify pocket-dependent gap amplitudes for both fully gapped and nodal superconducting states. Specifically, the nodal $d_{x^2-y^2}/d_{xy}$-wave pairing exhibits robust low-energy power-law behavior, distinct from a fully gapped pairing. Additionally, for the $s_{\pm}$-wave pairing, the detailed gap anisotropy on the $\beta$ pocket can be determined. Possible experimental implications are also discussed. Our results highlight the crucial role of multiorbital effects in shaping the Raman spectra and establish electronic Raman scattering as a powerful and symmetry-resolved probe for determining the superconducting gap in unconventional superconductors.
\end{abstract}

\maketitle

{\it Introduction}.
The recent discovery of superconductivity in the Ruddlesden-Popper bilayer nickelate La$_3$Ni$_2$O$_7$ (LNO) under pressure, exhibiting a transition temperature (T$_c$) nearly 80 K, has garnered significant research interest\cite{sun2023}. More recently, superconductivity at ambient pressure has been achieved in compressively strained LNO thin films grown on SrLaAlO$_4$ substrates, with a T$_c$ exceeding 40 K\cite{Ko2025,Zhou2025}. These advancements establish bilayer nickelates as the third high-T$_c$ family, besides cuprates and iron-based superconductors. In contrast to
its predecessors, the unique strongly-coupled bilayer structure through the inner apical oxygens results in a $d^{7.5}$ electronic configuration in Ni$^{2.5+}$, with both $d_{x^2-y^2}$ and $d_{z^2}$ orbitals dominating the low-energy physics~\cite{YaoDX,YZhang2023,Lechermann2023,KurokiFLEX,XWu,XJZhou2023,HHWen2023,Geisler20241,Geisler20242}.  In bulk LNO under pressure, superconductivity emerges abruptly following a structural transition from the \textit{Amam} phase at low pressure, which features density wave orders, to the \textit{Fmmm}/\textit{I4mmm} phase\cite{Liu2022,ShuLeiSDW,chen2024electronic,Kakoi2024,plokhikh2025,yashima2025,Khasanov2025,ZHAO2025,Ren2025}. Despite intensive studies, essential electronic structure for superconductivity remains unclear, and recent angle-resolved photoemission spectroscopy (ARPES) measurements on superconducting thin films have not reached a consensus~\cite{ShenAPERS,ChenAPERS}. 

The pairing mechanism and symmetry are subjects of ongoing theoretical debate. Various theoretical approaches have been employed, predicting different pairing symmetries. Calculations within weak-coupling frameworks, driven by spin fluctuations, predominantly favor an $s_\pm$-wave pairing ~\cite{Wang327prb,FangYang327prl,XWu,zhang2024structural,KurokiFLEX}, with $d_{xy}$-wave pairing also proposed in the presence of small crystal field splitting ~\cite{Lechermann2023,HHChenNC,GriffinFLEX}. In contrast, strong-coupling approaches emphasize the roles of interlayer exchange, Hund's coupling, and orbital hybridization, predicting interlayer $s$-wave ~\cite{lu2024interlayer,HYZhangtype2,WeiLi327prl} or in-plane $d_{x^2-y^2}$-wave pairing ~\cite{Jiang_2024,KuWeiprl,fan2023superconductivity}. Additionally, strong interlayer repulsion may drive interlayer inter-orbital $d_{x^2-y^2}$-wave pairing~\cite{JXLiFLEX,DiasRPA,zhan2025impact}. Consequently, experimentally identifying the pairing symmetry will constrain the possible pairing scenarios and help resolve debates regarding the underlying pairing mechanism.

As high pressure is necessary for superconductivity in bulk LNO, direct probes of the superconducting gap, such as ARPES and scanning tunneling microscopy (STM), remain extremely challenging. Indirect methods, such as point contact measurements, have led two groups to report inconsistent gap structures—$s$-wave and $d$-wave—in bulk LNO under pressure~\cite{cao2025direct,Guo2025}. While, preliminary ARPES~\cite{shen2025nodeless} and STM~\cite{fan2506superconducting} measurements on compressively strained thin films report a full gap along the diagonal direction and a nearly isotropic gap on three Fermi surface sheets. The discrepancies between bulk and thin-film materials may arise from differences in sample quality, disorder effects, and measurement methods, highlighting the need for a consistent probe applicable to both bulk and thin-film LNO. Electronic Raman scattering is a powerful tool to address this need, which can provide crucial insights into the superconducting gap due to its sensitivity to both gap magnitude and symmetry~\cite{DevereauxRMP}. Raman scattering involves the inelastic scattering of light when photons interact with electrons, phonons, or other elementary excitations. By manipulating the polarization vectors of incident and scattered photons, it is possible to identify the symmetry of pair-breaking excitations, which correlates with the symmetry of the superconducting gap. Given the multi-orbital nature of bilayer nickelates, the screening effect in the $A_{1g}$ symmetry channel is particularly intriguing, and different approximations of the Raman vertex can influence the Raman response results.
Therefore, Raman scattering offers a versatile and symmetry-resolved method to elucidate the superconducting order parameter in both bulk and thin-film LNO, thereby helping to resolve the ongoing debates.


In this paper, we study the Raman response in different channels for various pairing symmetries within a two-orbital bilayer model. We calculate Raman susceptibilities using both multiorbital and band-additive approaches, which exhibit qualitative consistency. By analyzing the Raman responses across different channels, we can distinguish different superconducting symmetries based on their low-energy behaviors and determine pocket-dependent gap amplitudes for both fully gapped and nodal superconducting states. Specifically, the nodal $d_{x^2-y^2}/d_{xy}$-wave pairing displays robust low-energy power-law behavior, in contrast to fully gapped pairings. Furthermore, for $s_{\pm}$-wave pairing, the detailed gap anisotropy on the $\beta$ pocket can be resolved. We also discuss the potential experimental implications of our findings.

\begin{figure}[t]
	\centering
	\includegraphics[width=0.35\textwidth]{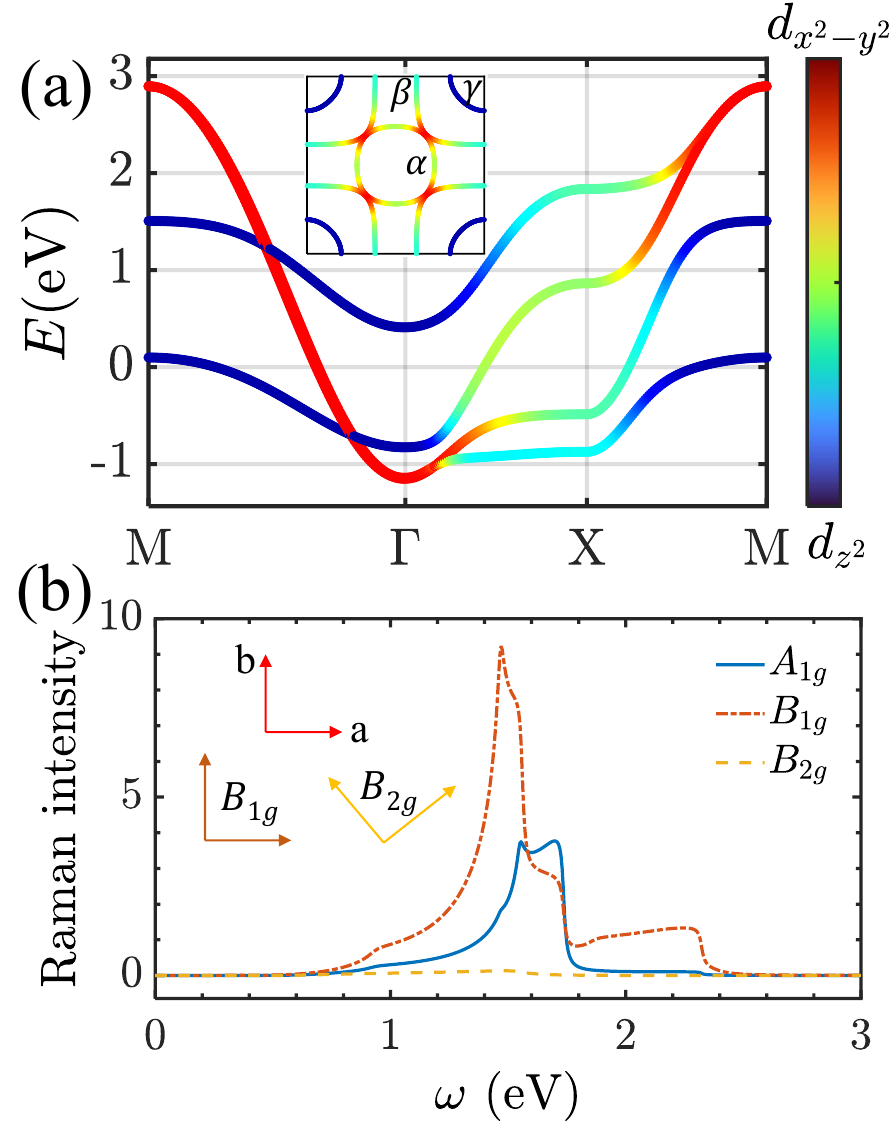}
	\caption{(a) Electronic structure of La$_3$Ni$_2$O$_7$ under pressure. The orbital-resolved band structure is shown, where colors represent the orbital contributions. The inset displays the Fermi surface obtained from the tight-binding model at filling $n=3$. (b) Normal-state Raman response above the interband transition energy scale. The inset illustrates the crystalline $a$ and $b$ axes, as well as the polarization geometries for the incident and scattered photons in the $B_{1g}$ and $B_{2g}$ channels.}
	\label{figTB327}
\end{figure}

{\it Model and pairing}.
We start with the low-energy electronic structures of bilayer nickelates La$_3$Ni$_2$O$_7$, which are dominated by Ni $d_{x^2-y^2}$ and $d_{z^2}$ orbitals according to theoretical calculations and experimental
measurements~\cite{YaoDX,YZhang2023,Lechermann2023,KurokiFLEX,XWu,XJZhou2023,HHWen2023}.
Therefore, the low-energy physics can be described by a bilayer two-orbital tight-binding (TB) 
Hamiltonian~\cite{YaoDX,YZhang2023,Lechermann2023,KurokiFLEX,XWu},
which reads,
\begin{equation}
	\mathcal{H}_0=\sum_{ij,\alpha\beta,\sigma}t_{\alpha\beta}^{ij}c_{i\alpha\sigma}^\dagger c_{j\beta\sigma}-\mu\sum_{i\alpha\sigma} c_{i\alpha\sigma}^\dagger c_{j\alpha\sigma}.
\end{equation}
Here $i,j=(m,l)$ labels the in-plane lattice site ($m$) and layer index ($l=t,b$), $\sigma$ labels spin, and $\alpha,\beta=x,z$ labels the Ni orbitals with $x$ denoting the $d_{x^2-y^2}$ and $z$ the $d_{z^2}$ orbital. $\mu$ is the chemical potential and the adopted hopping parameters are provided in Ref.~\cite{XWu}. 
In momentum space, the TB Hamiltonian can be written as $\mathcal{H}_{0}=\sum_{\mathbf{k}\sigma}\psi_{\mathbf{k}\sigma}^\dagger H_\mathbf{k}\psi_{\mathbf{k}\sigma}$ with basis operator $\psi_{\mathbf{k}\sigma}=[c_{tx\sigma}(\mathbf{k}),c_{tz\sigma}(\mathbf{k}),c_{bx\sigma}(\mathbf{k}),c_{bz\sigma}(\mathbf{k})]$. 
The corresponding band structure is shown in Fig.\ref{figTB327}(a), where $d_{x^2-y^2}$ and $d_{z^2}$ orbitals are represented by red and blue colors, respectively. The Fermi surfaces (FS) at the average filling $n=3$ per unit cell (1.5 per Ni atom) is illustrated in Fig.\ref{figTB327}(a), which contains three pockets: the $\alpha$ electron pocket arises from the interlayer bonding state of $d_{x^2-y^2}$ and $d_{z^2}$ orbital and the hole-like $\beta$ and $\gamma$ pockets originate mainly from the interlayer anti-bonding state of $d_{x^2-y^2}$ and bonding state of $d_{z^2}$ orbital, respectively.

In the superconducting state, we consider spin singlet pairing described by $4\times4$ order parameters $\Delta(\mathbf{k})$ in orbital space. The superconducting pairing Hamiltonian is given by
\begin{equation}
	\begin{aligned}
		\mathcal{H}_{\Delta} &= \sum_{\mathbf{k}l_1 l_2\alpha\beta} \Delta_{l_1\alpha, l_2\beta}(\mathbf{k})  c^{\dagger}_{\mathbf{k}l_1\alpha \uparrow } c^{\dagger}_{-\mathbf{k}l_2\beta\downarrow} + \mathrm{h.c.} \\
		&\approx \sum_{b} \Delta_{ b}(\mathbf{k}) c^{\dagger}_{b\uparrow }(\mathbf{k}) c^{\dagger}_{b\downarrow}(-\mathbf{k}) + \mathrm{h.c.} 
	\end{aligned}
\end{equation}


\noindent
In the second equation, we consider the dominant intraband pairing and interband pairings are neglected. The unitary matrix $\mathcal{U}(\mathbf{k})$ diagonalizes the TB Hamiltonian $\mathcal{H}_0$, $\mathcal{U}^\dagger(\mathbf{k}) H_{\mathbf{k}} \mathcal{U}(\mathbf{k}) = \tilde{H}_{\mathbf{k}}=\mathrm{diag}(\varepsilon_{1\mathbf{k}}, \dots, \varepsilon_{4\mathbf{k}})$, and implements the orbital-to-band transformation, $c_{l\alpha\sigma}(\mathbf{k}) = \sum_{b} \mathcal{U}_{l\alpha b}(\mathbf{k}) c_{b\sigma}(\mathbf{k})$,
and the intraband pairing gap is then given by 
$\Delta_{b}(\mathbf{k}) = \sum_{l_1 l_2 \alpha\beta} \mathcal{U}^{*}_{l_1\alpha b}(\mathbf{k}) \Delta_{l_1\alpha,l_2\beta}(\mathbf{k})  \mathcal{U}^{*}_{l_2\beta b}(-\mathbf{k})$ .

So far, various pairing symmetries have been proposed from theoretical calculations~\cite{Wang327prb,XWu,Lechermann2023,zhang2024structural,FangYang327prl,HHChenNC,Gao2025,DiasRPA,ryee2025optimal,KurokiFLEX,GriffinFLEX,JXLiFLEX,ushio2025theoretical,Wang327prb,EPC,QHWangpressure,le2025landscape,zhan2025impact,Cao2026,ZYLuDMFT,ryee2024quenched,WWUcDMFT1,Maier2026,WWUcDMFT2,Shen_2023,Schlömer2024,KurokiDMRG,WeiLi327prl,JLChenPRB,Wang327prb,Luo2024,Xue_2024,lu2024interlayer,HYZhangtype2,Jiang_2024,CJWUSBMF2,HYZhangtype22,fan2023superconductivity,Jiang_2024,KuWeiprl,oh2025high,liu2025variational}, such as $s_{\pm}$-, $d_{x^2-y^2}$- and $d_{xy}$-wave pairings, but no consensus has been
achieved
in experiments. These pairing states can be expressed in the orbital and layer space. 
In the following, we use $\bm{\sigma}$ and $\bm{\tau}$ to denote the Pauli matrices in the layer and orbital spaces, respectively and $\tau_{\pm}=(\tau_0\pm\tau_3)/2$.

\begin{figure}[t]
	\centering
	\includegraphics[width=0.35\textwidth]{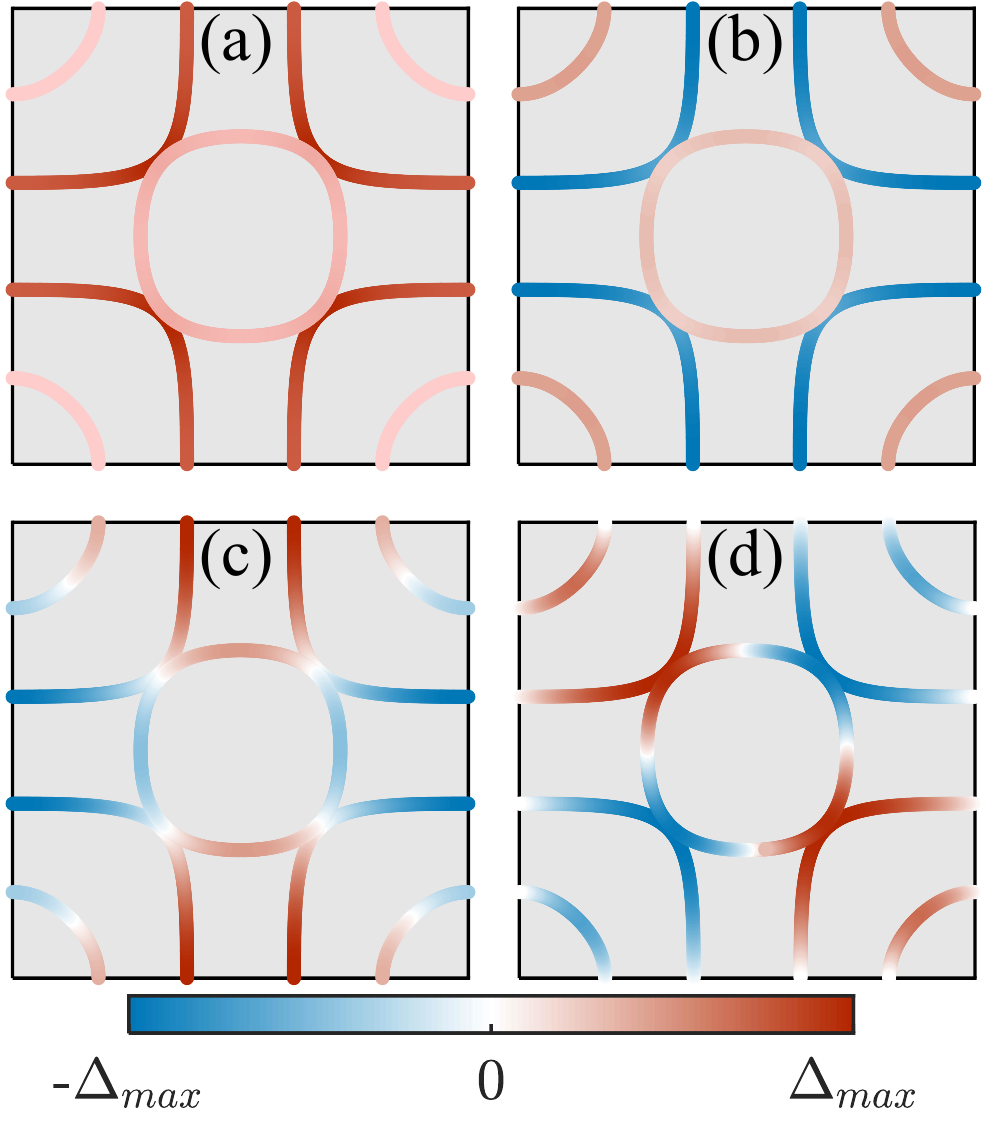}
	\caption{Superconducting gap on the Fermi surface for four representative pairing states: (a) intralayer-dominated $s_{++}$, (b) interlayer-dominated $s_{\pm}$, (c) in-plane $d_{x^2-y^2}$ pairing, and (d) in-plane $d_{xy}$ pairing.}
	\label{figSCgap}
\end{figure}

We consider four representative pairing states, with their gap functions shown in the Fig.~\ref{figSCgap}. Their characteristics and pairing forms in the orbital space are listed in the following:

\begin{itemize}
	
	\item[(a)] {\it Intralayer-dominated $s_{++}$-wave pairing}. The pairing gaps are in phase on three pockets, shown in Fig.~\ref{figSCgap}(a). By adopting the pairing term $\Delta_{s_{++}} = \Delta_{0}(\sigma_0 - \sigma_1/2)(\tau_{0}-\tau_{-}/2) $, the superconducting gaps are pocket dependent due to different orbital characters on three pockets. In addition, the gap amplitude on the $\beta$ pocket is anisotropic.
	
	\item[(b)] {\it Interlayer-dominated $s_{\pm}$-wave pairing}. As shown in Fig.~\ref{figSCgap}(b), the phase of pairing on the antibonding pocket is opposite to those on the bonding pockets, stemming from the interlayer pairing. By taking $\Delta_{s_{\pm}} = \Delta_{0}(\sigma_1 - \sigma_0 /2)(\tau_{0}-\tau_{+}/2)$, the superconducting gaps are pocket dependent and gap anisotropy occurs on the $\beta$ pocket.
	
	\item[(c)] {\it Intralayer $d_{x^2-y^2}$-wave pairing}. As shown in Fig.~\ref{figSCgap}(c), typical line nodes occur along the diagonal direction. The intraorbital nearest-neighbor (NN) pairing can be written as $\Delta^{d_{x^2-y^2}}_{\mathrm{IP}} = \frac{1}{2}\Delta_0 \sigma_0\tau_0(\cos{k_x} - \cos{k_y})$ and the gap amplitudes are pocket dependent due to the form factors. The $d_{x^2-y^2}$-wave pairing can also be in the interlayer interorbital channel~\cite{JXLiFLEX,DiasRPA,zhan2025impact}, resulting orbital-dependent gap size.  
	
	\item[(d)] {\it Intralayer intraorbital $d_{xy}$-wave pairing}. Typical line nodes along $k_{x,y}=0,\pi$ as shown in Fig.~\ref{figSCgap}(d). By taking the next nearest-neighbor form factors, the pairing form is $\Delta^{d_{xy}}_{\mathrm{IP}} = \Delta_0 \sigma_0\tau_0  \sin{k_x}\sin{k_y}$ and the gap sizes are weakly dependent on pockets. 
\end{itemize}
In the superconducting state, the total Hamiltonian can be written as $\mathcal{H}_{\text{BdG}}=\mathcal{H}_0+\mathcal{H}_{\Delta}$, which we use to study the Raman response for different pairing states.  


{\it Formalism of Electronic Raman scattering in multiorbital systems}. Raman process is an inelastic scattering of polarized light by elementary excitations in materials~\cite{DevereauxRMP}.
The cross section of the scattered light is proportional to the imaginary part of
the channel-dependent Raman susceptibility $\chi^{\gamma\gamma}$.
Due to the complexity of photon–electron interaction vertices, the effective mass approximation is widely employed to construct the Raman vertex in nonresonant electronic Raman scattering~\cite{DevereauxRMP}.
In general multiorbital systems, within the effective mass approximation, the Raman vertex describing effective Raman charge fluctuations in the $\gamma$ channel can be expressed in the orbital space as
\begin{equation}
	R_{\mathbf{k}}^{\gamma}= \begin{cases}\left(\frac{\partial^2}{\partial k_x^2} + \frac{\partial^2}{\partial k_y^2}\right)H_{\bf k} , & \gamma=A_{1 g}, \\ \left(\frac{\partial^2}{\partial k_x^2} - \frac{\partial^2}{\partial k_y^2}\right)H_{\bf k} , & \gamma=B_{1 g}, \\ 2\frac{\partial^2 }{\partial k_x \partial k_y} H_{\mathbf{k}}, & \gamma=B_{2 g}.\end{cases}
\end{equation}
\noindent 
Here $ H_{\mathbf{k}}$ is the single-particle Hamiltonian in the momentum space and $A_{1g},B_{1g},B_{2 g}$ are three typical Raman channels in experiments. From these expressions, we observe that $R_{\mathbf{k}}^{A_{1g},B_{1g}}$ is proportional to the NN hopping, while $R_{\mathbf{k}}^{B_{2g}}$ is proportional to the next NN hopping. 
Through a unitary transformation, the Raman vertex in the band space reads
$R_{\mathbf{k},B}^\gamma = \mathcal{U}^\dagger(\mathbf{k}) R^\gamma_{\mathbf{k}} \mathcal{U}(\mathbf{k})$.
 The bare Raman susceptibility of the $\gamma$ channel in the superconducting state is defined as 
\begin{equation}\label{eq:sus1}
	\begin{aligned}
		& \chi^{\gamma}_{\mathrm{MO}}\left( i \omega_n\right) 
		= \frac{-1}{\beta V} \sum_{\mathbf{k}, \omega_m}   \operatorname{Tr} \left[ G\left(\mathbf{k}, i \omega_m\right) \tilde{R}^{\gamma}_{ \mathbf{k}}  G\left(\mathbf{k}, i \omega_m+i \omega_n\right)  \tilde{R}^{\gamma}_{ \mathbf{k}} \right]. 
	\end{aligned}
\end{equation}
Here $G\left(\mathbf{k}, i \omega_m\right)=[i \omega_m-H_{\text{BdG}}(\mathbf{k})]^{-1}$ is  Gorkov Green function and $\tilde{R}_{\mathbf{k}}^{\gamma}=R_{\mathbf{k}}^{\gamma} \otimes\rho_3$ is the Raman vertex in superconducting state with $\bm{\rho}$ being Pauli matrices in Nambu space. 
Transforming Green functions and Raman vertices in the above equations into the band basis, we will find that contributions from both intraband and interband processes are incorporated. 
The Raman response $\chi^{\gamma}(\omega)$ is calculated as the imaginary part of $\chi^{\gamma}\left( \omega + i\delta \right)$ after the analytical continuation $i\omega_n \to \omega + i\delta$.


The Raman scattering in superconductors can be also studied in the band-additive approximation, where the total response is the addition of the response contribution from each band separately\cite{devereaux96,boyd09,sauer82}. The total Raman response $\chi_{\rm BA}^{\gamma}$ can be written as
\begin{eqnarray}
	\label{Rasauer1}
	\chi_{\rm BA}^{\gamma}(i\omega_n) &=& \sum_{b} \chi_{\gamma \gamma, b}(i\omega_n),\\
	\chi_{\gamma \gamma,b}(i\omega_n) &=&
		\sum_{{\bf k}}
		\gamma_{b\mathbf{k}}^2 \, \frac{\Delta^2_{b\mathbf{k}}}{E^2_{b\mathbf{k}}}
		\tanh \frac{E_{b\mathbf{k}}}{2T}
		\nonumber\\ 
		 &&\times \left[ \frac{1}{i\omega_n + 2E_{b\mathbf{k}}} - 
		\frac{1}{i\omega_n - 2E_{b\mathbf{k}}}\right] \, ,
		\label{eq:fullRaman_n}
\end{eqnarray}
where the energy of quasiparticles in the superconducting state reads $E_{b\mathbf{k}}=\sqrt{\varepsilon^2_{b\mathbf{k}}+\Delta^2_{b\mathbf{k}}}$ and $\chi_{\gamma \gamma, b}$ is the bare Raman susceptibility calculated for a given band $b$. The corresponding Raman vertices in the band space are defined as,
\begin{equation}
	\gamma_{b}(\mathbf{k})= \begin{cases}\frac{\partial^2 \varepsilon_{b\mathbf{k}}}{\partial k_x^2}+\frac{\partial^2 \varepsilon_{b\mathbf{k}}}{\partial k_y^2}, & A_{1 g}, \\ \frac{\partial^2 \varepsilon_{b\mathbf{k}}}{\partial k_x^2}-\frac{\partial^2 \varepsilon_{b\mathbf{k}}}{\partial k_y^2}, & B_{1 g}, \\ 2\frac{\partial^2 \varepsilon_{b\mathbf{k}}}{\partial k_x \partial k_y}, & B_{2 g}.\end{cases}
\end{equation}
These Raman vertices directly depend on the curvatures of the electronic bands and are slightly different from those in the orbital basis, which will result in different Raman responses in the two approximations.

The Raman response of different channels in the normal state is displayed in the Fig. \ref{figTB327}(b).
Only inter-band transitions can contribute to the Raman scattering, which occurs above $\omega \sim 0.8 $ eV.
The Raman responses in $A_{1g}$ and $B_{1g}$ channels peak in the energy range between 1.4 eV to 1.8 eV, which is attributed to the transition between bonding and antibonding $d_{z^2}$ bands around the M point and between states around the X point. 
In the superconducting state, pair breaking features appear at low energies, corresponding to the magnitude of the pairing gap, which is well  below the inter-band transitions.
The low-energy pair-breaking characteristics in the Raman response will be used to distinguish different pairing symmetries.

{\it Electronic Raman signals for different pairing symmetries}. We further study the Raman response for different pairing symmetries in bilayer nickelates. In the following, we will show Raman susceptibilities in both multiorbital (MO) and band-additive (BA) formalism to reveal intrinsic low-energy features. The Raman response is insensitive to the sign change of the superconducting gap on different bands and therefore in principle cannot effectively distinguish between the $s_{++}$- and $s_{\pm}$-wave pairings. However, it is sensitive to both gap symmetry and gap anisotropy, especially nodal gap. 

\begin{figure}[t]
	\centering
	\includegraphics[width=0.49\textwidth]{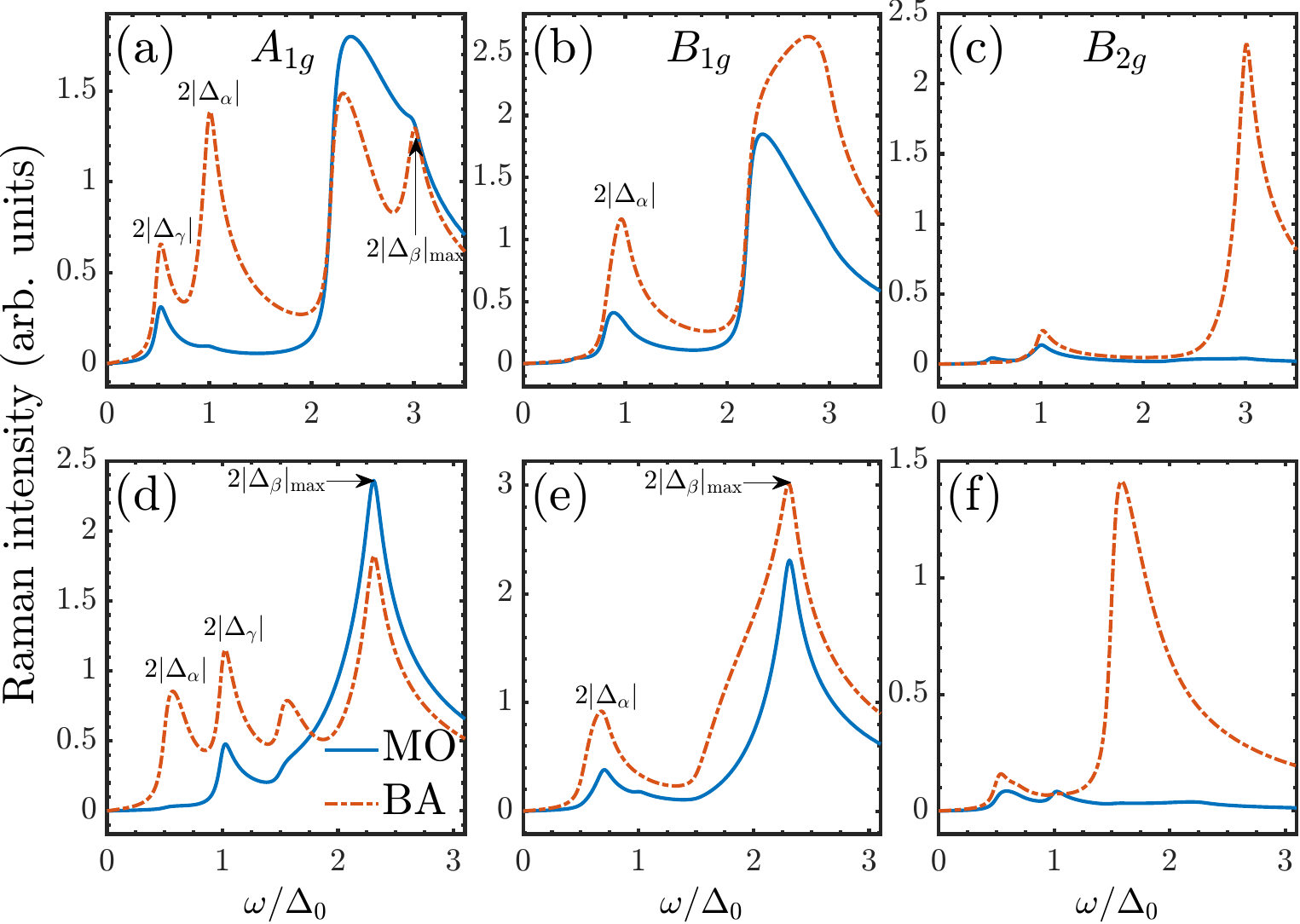}
	\caption{
	The Raman responses in the $A_{1g}$ (a,d), $B_{1g}$ (b,e), and $B_{2g}$ (c,f) channels 
	for the $s_{\pm}/s_{++}$-wave pairing with the gap maximum on $\beta$ pocket along the diagonal direction (a-c) and around X points (d-f), corresponding to Figs.~2(a) and 2(b). 
	The blue solid curves represent the full multiorbital (MO) calculations, while the orange dotted curves 
	denote the band-additive (BA) Raman responses. The energies of the dominant peak positions are indicated.
	}
	\label{figR1}
\end{figure}

In the $s_{++}/s_{\pm}$ pairing, uniform gaps on three pockets will result in a pair-breaking peak at twice the superconducting gap on Fermi surface. But theoretical calculations suggest that electronic interaction will drive pocket-dependent gaps with some anisotropy, especially on the $\beta$ pocket~\cite{EPC,KurokiFLEX,JXLiFLEX}. For the anisotropic $s_{++}/s_{\pm}$-wave pairing, we consider two cases:  the $\beta$ pocket exhibits gap maximum along the diagonal direction or around saddle points X/Y. By taking appropriate pairing parameters, the gaps in these two cases are displayed in Fig.\ref{figSCgap} (a) and (b), where the gap on the $\beta$ pocket is larger than those on the $\alpha$ and $\gamma$.

In the former case, the Raman responses in the three channels are illustrated in Fig.~\ref{figR1}(a)-(c) and low-energy Raman signals almost vanish until certain energy scale where pair-breaking features occur. 
Motivated by recent ARPES~\cite{shen2025nodeless} and STM~\cite{fan2506superconducting} measurements on compressively strained superconducting thin films, we adopt a superconducting gap amplitude of $\Delta_0 = 20$~meV and a broadening of  $\delta=1$~meV  in our calculations,generating consistent gap amplitudes with experiments.
In the $A_{1g}$ channel, multiorbital Raman susceptibility $\chi^{\gamma}_{\text{MO}}$ exhibits a weak pair-breaking peak around twice the smallest gap on the $\gamma$ pocket ($2|\Delta_{\gamma}|=0.5\Delta_0$) and a broader peak around the energy scale ($2|\Delta_{\beta}|_{\text{min}}=2.2\Delta_0$) corresponding to the gap on the $\beta$ pocket with an additional kink feature at $2|\Delta_{\beta}|_{\text{max}}=3\Delta_0$. The broadening of the latter peak originates from the gap anisotropy on the $\beta$ pocket, where the locations of the maximum gap, maximum Raman vertices and the density of states (DOS) maxima do not coincide. In contrast, the band-additive susceptibility $\chi^{\gamma}_{\text{BA}}$ produces  two $\beta$ pocket related peaks at $2.2\Delta_0$, $3\Delta_0$, coinciding with energies of peak and kink features in $\chi^{\gamma}_{\text{MO}}$, and an additional peak around $2|\Delta_{\alpha}|=\Delta_0$. The discrepancy of the Raman intensity and appearance of the peak feature at $2|\Delta_{\alpha}|$ between the two approaches arises from the distinct structures of the Raman vertices.
In the $B_{1g}$ channel, the Raman vertices exhibit nodes along the diagonal direction and are strongly suppressed on the $\gamma$ pocket. Consequently, the $\gamma$ pocket will not contribute with a pair-breaking signal and the first peak appears at larger energy around $2|\Delta_{\alpha}|$. In addition, the signal at $2|\Delta_{\beta}|_{\text{max}}$ is suppressed and both approaches give rise to a broad peak at energy slightly smaller than $2|\Delta_{\beta}|_{\text{max}}$.
In the $B_{2g}$ channel, the Raman vertices exhibit nodes for $k_{x,y}=0,\pi$ lines and are weak on both $\alpha$ and $\gamma$ pockets. The Raman response $\chi^{\gamma}_{\text{MO}}$ is vanishingly small compared to those in the $A_{1g}$ and $B_{1g}$ channels due to weak next NN hopping and the weak peak around $2|\Delta_{\alpha}|$ is still visible. By contrast, the $\chi^{\gamma}_{\text{BA}}$ shows an additional sharp peak around $2|\Delta_{\beta}|_{\text{max}}$, which is attributed to enhanced Raman vertices along the diagonal direction.

In the latter case, the gap maximum of the $\beta$ pocket appears around X points and the gap size on the $\gamma$ pocket is larger than the $\alpha$ pocket. The corresponding Raman response is shown in Fig.~\ref{figR1}(d)-(f). In the $A_{1g}$ channel, $\chi^{\gamma}_{\text{MO}}$ exhibit one peak at $2|\Delta_{\gamma}|=\Delta_0$ and another sharp peak around $2|\Delta_{\beta}|_{\text{max}}=2.3\Delta_0$. This peak originates from the coincidence of the positions of the gap maximum and the DOS maximum on the $\beta$ pocket. However, $\chi^{\gamma}_{\text{BA}}$ displays four pair-breaking features: two smaller peaks at $0.6\Delta_0$ and $\Delta_0$ come from the $\alpha$ and $\gamma$ pockets, and two larger ones at $1.5\Delta_0$ and $2.3\Delta_0$ are associated with the anisotropic gap on the $\beta$ pocket. In the $B_{1g}$ channel, both $\chi^{\gamma}_{\text{MO}}$  and $\chi^{\gamma}_{\text{BA}}$  show similar pair-breaking features around $2\Delta_{\alpha}$ and $2|\Delta_{\beta}|_{\text{max}}$. In the $B_{2g}$ channel, $\chi^{\gamma}_{\text{MO}}$ shows
peaks around $2\Delta_{\alpha}$ and $2\Delta_{\gamma}$ although its intensity is low. While $\chi^{\gamma}_{\text{BA}}$ exhibits an additional sharp peak at  $2|\Delta_{\beta}|_{\text{min}}$. Therefore, the Raman response under different channels can detect both  pocket dependent gap sizes and gap anisotropy for the $s_{++}/s_{\pm}$ pairing.

\begin{figure}[t]
	\centering
	\includegraphics[width=0.49\textwidth]{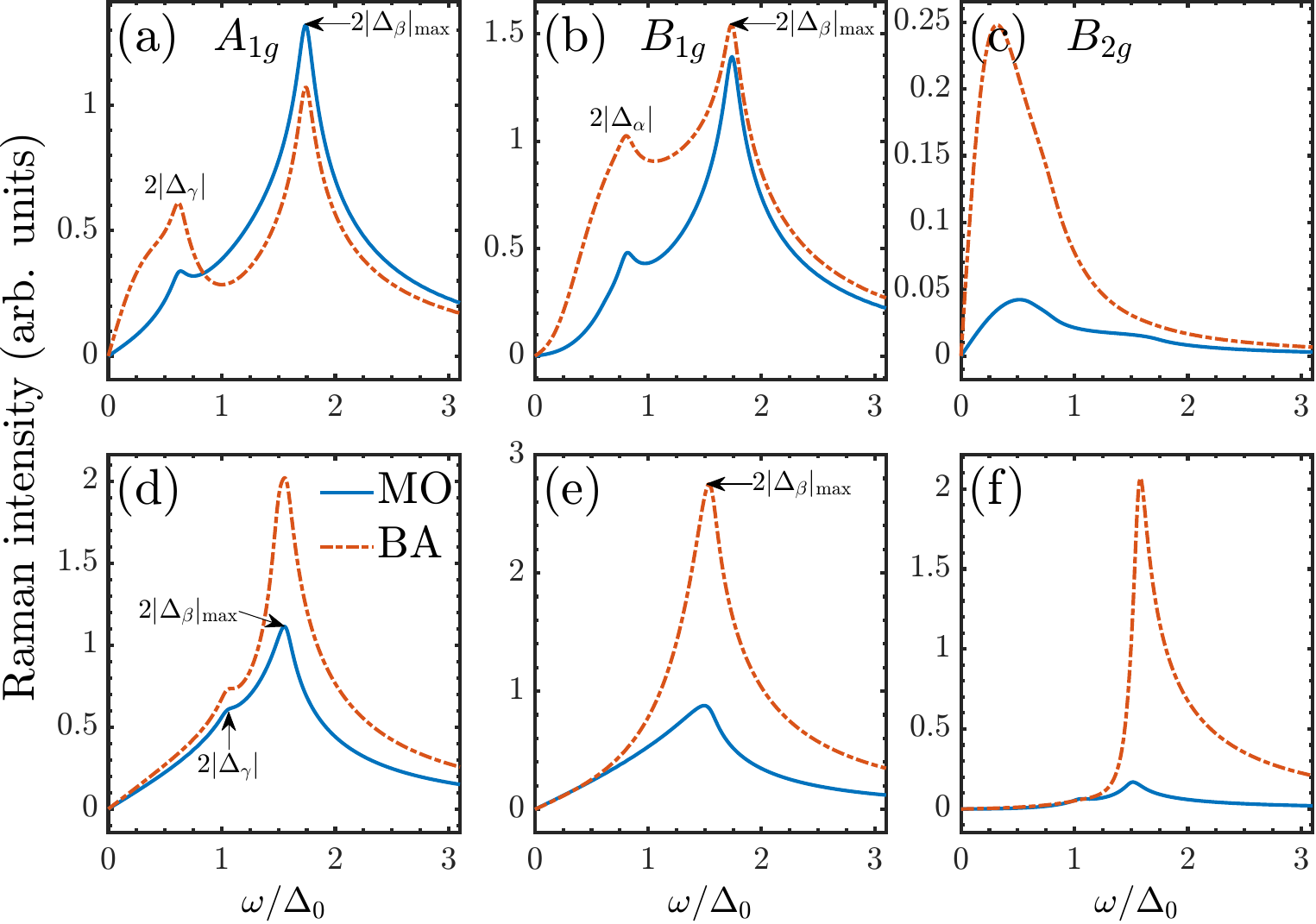}
	\caption{
	The Raman responses in the $A_{1g}$ (a,d), $B_{1g}$ (b,e), and $B_{2g}$ (c,f) channels for in-plane 
	$d_{x^2 - y^2}$ (a–c) and $d_{xy}$ (d–f) pairing states corresponding to Figs.~2(c) and 2(d). 
	The blue solid curves represent the full multiorbital (MO) calculations, while the orange dotted curves denote 
	the band-additive (BA) Raman responses. The energies of the dominant peak positions are indicated.
	}
	\label{figR2}
\end{figure}

Moreover, we investigate the Raman responses for the two nodal 
$d_{x^2-y^2}$- and $d_{xy}$-wave pairings, as shown in Fig.~\ref{figR2}. 
The results obtained from the two approaches are qualitatively consistent, 
and the low-energy power-law behaviors in different symmetry channels 
indicate the presence of nodes in the superconducting gaps. 
In the $A_{1g}$ channel, $\chi^{\gamma}_{\text{MO}}$ exhibits a pronounced peak 
around $2|\Delta_{\beta}|_{\text{max}} = 1.7\Delta_0 / 1.5\Delta_0$, 
together with a smaller kink near $2|\Delta_{\gamma}| = 0.6\Delta_0 / 1\Delta_0$ 
for the $d_{x^2-y^2}/d_{xy}$-wave pairings. 
The contribution from the $\alpha$ pocket is minor due to the suppressed Raman vertices, 
similar to the $s$-wave case. Both pairings exhibit linear low-energy behaviors 
in the $A_{1g}$ channel, reflecting the presence of gap nodes.
In contrast, in the $B_{1g}$ channel, the Raman vertices on the $\alpha$ pocket 
are enhanced but become suppressed on the $\gamma$ pocket. 
Consequently, for the $d_{x^2-y^2}$ pairing, the $B_{1g}$ Raman response 
shows a dominant peak near $2\Delta_{\beta}$ and a weaker feature 
around $2|\Delta_{\alpha}| = 0.8\Delta_0$. 
The low-energy Raman intensity varies as $\omega^3$ 
because the nodes of the gap coincide with the nodes of the Raman vertices. 
For the $d_{xy}$ pairing state, a single pair-breaking peak appears at 
$2|\Delta_{\beta}|_{\text{max}}$, as the maxima of the Raman vertices 
coincide with the nodal directions of the $d_{xy}$-wave gap on the $\alpha$ pocket. 
In this case, the low-energy Raman intensity varies linearly with $\omega$.
In the $B_{2g}$ channel, $\chi^{\gamma}_{\text{BA}}$ exhibits a peak 
around $2|\Delta_{\gamma}|$ and $2|\Delta_{\beta,\alpha}|_{\text{max}}$ 
for the $d_{x^2-y^2}$- and $d_{xy}$-wave pairings, respectively. 
This behavior arises from the large Raman vertices along the diagonal direction 
in the $B_{2g}$ channel within the band-additive approach. 
Although the overall intensity of $\chi^{\gamma}_{\text{MO}}$ is much weaker, 
its general features remain consistent with those of $\chi^{\gamma}_{\text{BA}}$.


{\it Discussion and conclusion}. The Raman response is strongly correlated with the Raman vertices and the symmetry of the superconducting gaps. In our theoretical calculations, we adopted Raman vertices in three channels within different approximations and find that most of prominent features in two approaches are consistent. Theoretical calculations suggest that the gap on the $\beta$ pocket is highly anisotropic and accidental nodes can appear in the $s_{\pm}$-wave pairing~\cite{EPC,JXLiFLEX}, which will lead to some low-energy power-law behaviors in the Raman signal. Although the direct comparison between experimental and theoretical Raman response may be extremely challenging, our calculations can help to interpret future Raman results in experiments and identify the gap scales and symmetry. Recently, preliminary ARPES~\cite{shen2025nodeless} and STM~\cite{fan2506superconducting} measurements on compressively strained thin films reveal a full gap along the diagonal direction and an almost isotropic gap across the three Fermi surface sheets. While, two studies on Andreev reflection spectroscopy in bulk La$_3$Ni$_2$O$_7$ under pressure~\cite{cao2025direct,Guo2025} have reported conflicting evidence about the gap symmetry: one is $d$-wave and the other is $s$-wave with two full gaps. The future high-pressure Raman measurements can be done in both bulk and thin-film La$_3$Ni$_2$O$_7$ under different symmetry channels. According to the distinctive characteristics of different pairing symmetries, it can provide unambiguous evidence to distinguish between candidate pairing symmetries and help to resolve the ongoing debate on the pairing mechanism in this emerging family of high-$T_c$ superconductors.


In conclusion, we investigate the Raman response of bilayer nickelates in different channels for various pairing symmetries within a two-orbital bilayer model. Raman susceptibilities are performed in multiorbital and band-additive formalisms and they are qualitatively consistent. For the $s_{++}/s_{\pm}$-wave pairing, gap amplitudes on different pockets can be determined by Raman responses across different symmetry channels. In addition, the detailed gap anisotropy on the $\beta$ pocket can be identified by comparing $A_{1g}$ and $B_{1g}$ Raman signals. 
In analogy to high-$T_c$ cuprates, the Raman responses for the $d$-wave pairings exhibit characteristic low-energy power laws enforced by symmetry: a $\omega^3$ dependence for the $d_{x^2-y^2}$ pairing in the $B_{1g}$ channel and for the $d_{xy}$ pairing in the $B_{2g}$ channel, while other channels show linear behavior. These low-energy scaling behaviors are symmetry-determined and remain robust against the detailed form of the superconducting gap, thus serving as a clear fingerprint of $d$-wave pairing in the Raman spectra. 
Although theoretical studies have suggested	the possibility of s±pairing with accidental nodes~\cite{Wang327prb,EPC}, which may lead to similar low-energy power-law behavior as $d$-wave pairing, a combined analysis of the low-energy scaling behavior and pair-breaking features on different pockets in distinct symmetry channels can still provide strong evidence for distinguishing these pairing states.
Moreover, we find a noticeable orbital dependence of Raman vertices in the multiorbital scenario, in contrast to the band-additive scenario.  
The present analysis can be straightforwardly generalized to the case without a 
$\gamma$ pocket (see the Supplementary Material), further supporting electronic Raman scattering as a powerful probe for systems with different Fermi-surface topologies.
Our results highlight the critical role of multiorbital effects in shaping the Raman response and establish Raman scattering as a powerful probe of pairing symmetry in bilayer nickelates, which can help to resolve the ongoing debate on pairing mechanism.

{\it Acknowledgments}. We acknowledge the supports by National Natural Science Foundation of China (No. 12494594, No.11920101005, No. 11888101, No. 12047503, No. 12322405, No. 12104450), the Ministry of Science and Technology (Grant No. 2022YFA1403901), and  the New Cornerstone Investigator Program. X.W. is supported by the National Key R\&D Program of China (Grant No. 2023YFA1407300) and the National Natural Science Foundation of China (Grants No. 12574151, 12447103 and 12447101). A.G. and M.B. acknowledge the
Max-Planck-Institute for Solid State Research in Stuttgart for hospitality and financial support. A.P.S. is funded by the Deutsche Forschungsgemeinschaft (DFG, German Research Foundation) - TRR 360 - 492547816.

\bibliography{ref}
\bibliographystyle{apsrev4-1}

\appendix
\setcounter{equation}{0}  
\setcounter{figure}{0}  
\renewcommand{\thefigure}{A\arabic{figure}}
\renewcommand{\theequation}{A\arabic{equation}}
\begin{widetext}
	\section{Raman response for a Fermi surface without the $\gamma$ pocket}
	\begin{figure}[b]
		\centerline{\includegraphics[width=0.7\textwidth]{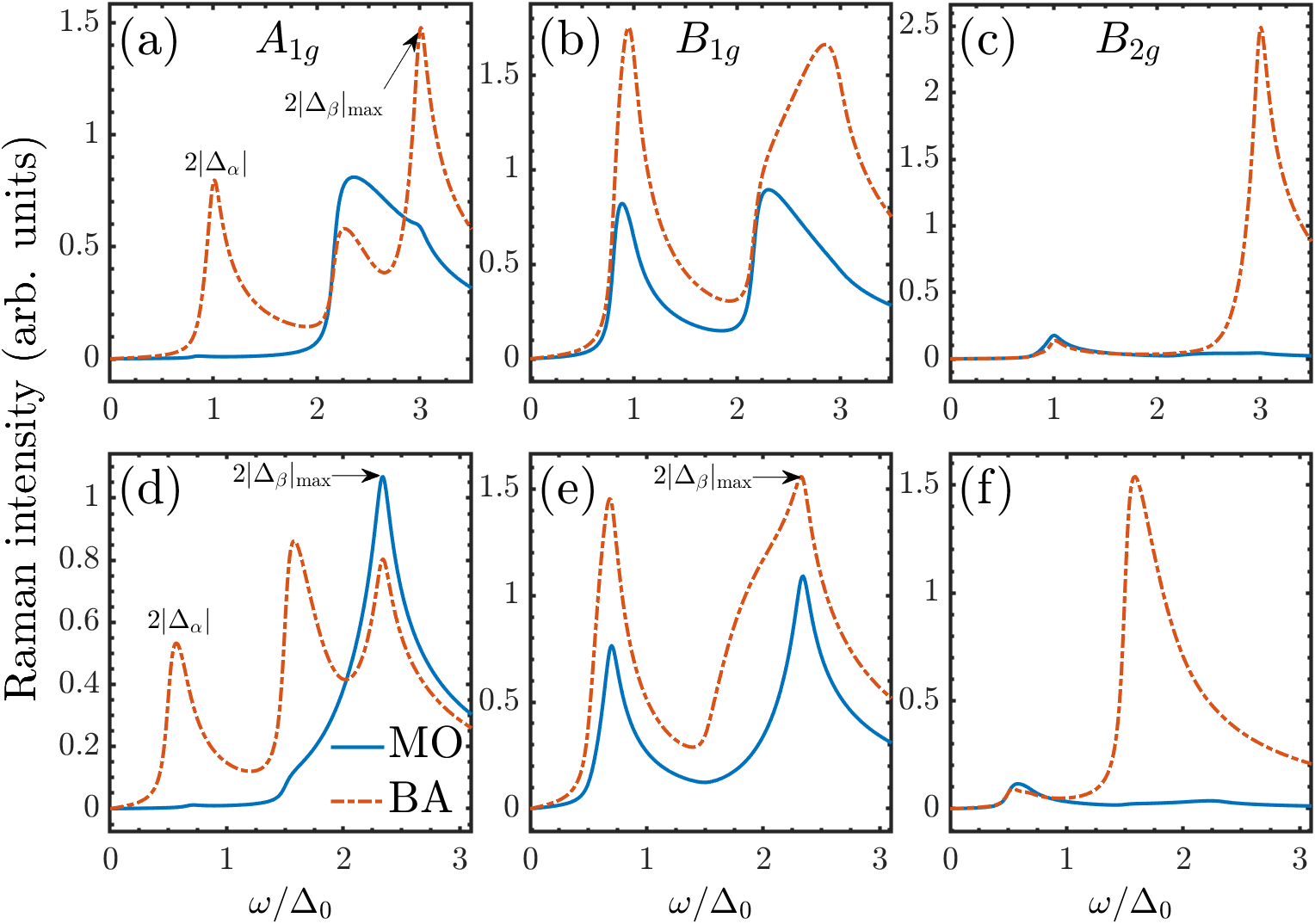}}
		\caption{
			Raman responses in the $A_{1g}$ (a,d), $B_{1g}$ (b,e), and $B_{2g}$ (c,f) channels for the $s_{++}$ (a--c) and $s_{\pm}$ (d--f) pairing states, using the same gap functions as in the main text but at a filling of $n = 1.7$ per site, where the $\gamma$ pocket is absent.
			The blue solid curves represent the full multiorbital (MO) calculations, while the orange dotted curves denote the band-additive (BA) Raman responses.
			The energies of the dominant pair-breaking peak positions are indicated.
			\label{fig000}}
	\end{figure}
	
	To investigate the Raman response for different Fermi-surface topologies, in particular the case without the $\gamma$ pocket whose existence remains under debate, we tune the Fermi level to an electron-doped regime in which 0.4 electrons are added and the $\gamma$ band becomes fully occupied. The corresponding Raman responses for the $s$-wave and $d$-wave pairing states considered in the main text are shown in Fig.~\ref{fig000} and Fig.~\ref{fig001}, respectively.
	
	In the absence of the $\gamma$ pocket, the pair-breaking features associated with this pocket naturally disappear. Nevertheless, the characteristic low-energy power-law scaling behaviors as well as the pair-breaking peaks originating from the remaining $\alpha$ and $\beta$ pockets persist. These results demonstrate that the key qualitative features of the Raman response are robust against changes in the Fermi-surface topology.
	
	Our analysis for different Fermi-surface topologies further supports electronic Raman scattering as a powerful and versatile probe for detecting the pairing symmetry in bilayer nickelates.

	\begin{figure}[t]
		\centerline{\includegraphics[width=0.7\textwidth]{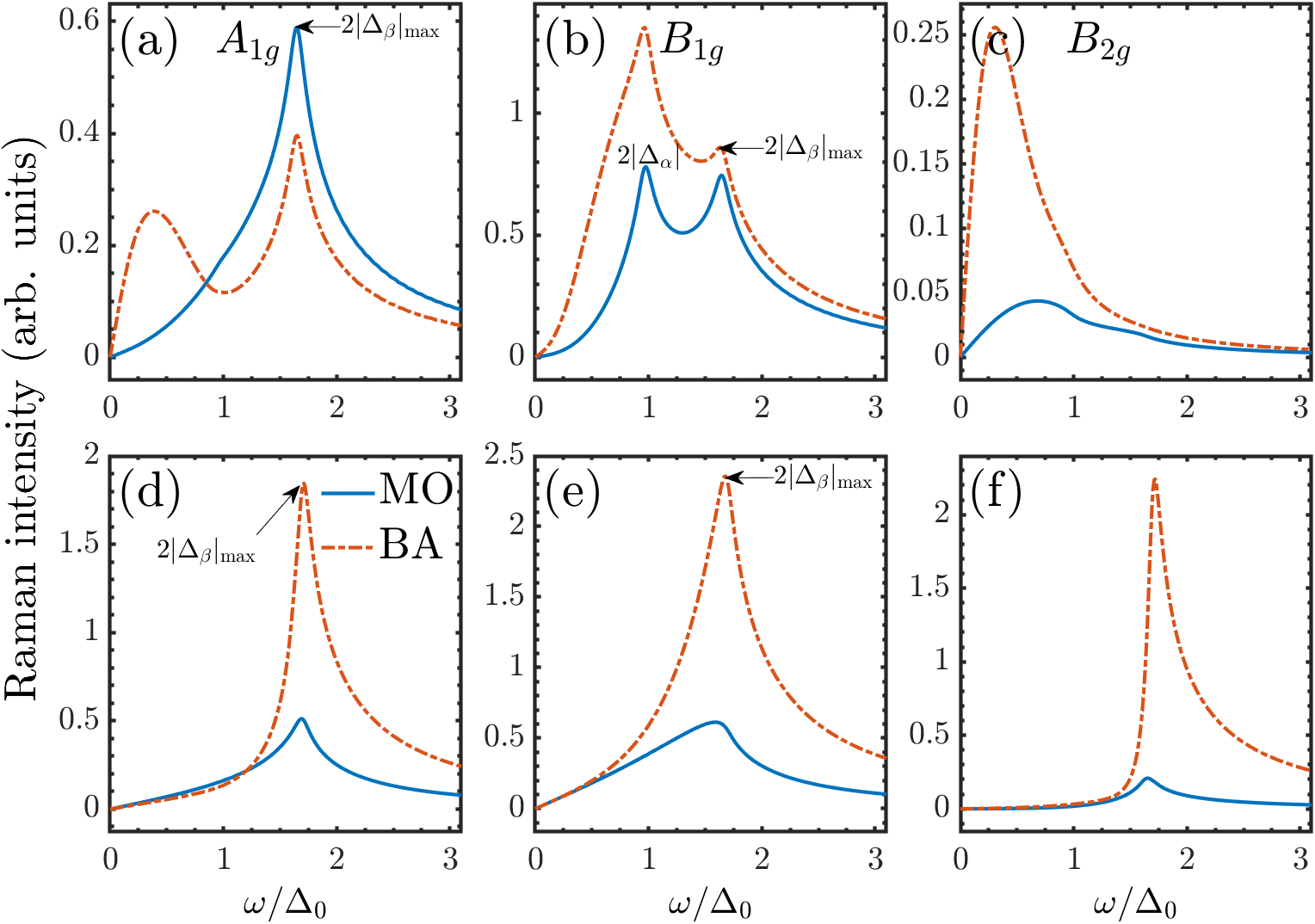}}
		\caption{
			Raman responses in the $A_{1g}$ (a,d), $B_{1g}$ (b,e), and $B_{2g}$ (c,f) channels for the $d_{x^2-y^2}$ (a--c) and $d_{xy}$ (d--f) pairing states, using the same gap functions as in the main text but at a filling of $n = 1.7$ per site, where the $\gamma$ pocket is absent.
			The blue solid curves represent the full multiorbital (MO) calculations, while the orange dotted curves denote the band-additive (BA) Raman responses.
			The energies of the dominant pair-breaking peak positions are indicated.
			\label{fig001}}
	\end{figure}
	
\end{widetext}

\end{document}